\documentclass{article}
\usepackage[normalem]{ulem}
\usepackage{graphicx}
\usepackage{latexsym,amssymb,amsmath,amssymb,xspace,theorem}
\usepackage{multicol}
\usepackage[latin1]{inputenc}
\usepackage[T1]{fontenc}
\usepackage{epic,eepic}
\usepackage{subfigure}
\usepackage{color}
\usepackage{epsfig}
\theoremstyle{plain} \theoremheaderfont{\scshape}

\newtheorem{Thm}{\bf Theorem}
\newtheorem{Lem}[Thm]{\bf Lemma}
\newtheorem{Clm}{Claim}[Thm]

\newtheorem{ClaimSansNum}{Claim}[Thm]
\newtheorem{Conj}[Thm]{{\bf Conjecture}}

\newtheorem{Prop}[Thm]{\bf Proposition}

{\theorembodyfont{\rmfamily}
 
 \newtheorem{Rem}[Thm]{\bf Remark}

}

\newenvironment{Prf}{{\bf \noindent Proof } }{\hfill$\square$\\}
\newenvironment{PrfClaim}{{\bf Proof }}{{\hfill\tiny{$\square$\\}}}

\newcommand{\ignore}[1]{}

\newcommand{\cqfd}{\unskip\kern 6pt\penalty 500
\raise -2pt\hbox{\vrule\vbox to 10pt{\hrule width 4pt
\vfill\hrule}\vrule}\par}

\graphicspath{./}

\input{epsf.sty}
\begin{document}
\title{M\'acajov\'a and \v{S}koviera Conjecture on Cubic Graphs.}
\author{J.L. Fouquet and J.M. Vanherpe\\
L.I.F.O., Facult\'e des Sciences, B.P. 6759 \\
Universit\'e d'Orl\'eans, 45067 Orl\'eans Cedex 2, FR
}
\maketitle
\begin{abstract}
A conjecture of M\'a\u{c}ajov\'a and \u{S}koviera
\cite{MacSko07} asserts  that every bridgeless cubic graph has two
perfect matchings whose intersection does not contain any odd edge
cut. We prove this conjecture for graphs with few vertices and we
give a stronger result for traceable graphs.\\
{\bf keywords~:}Cubic graph;  Edge-partition; Traceable graphs
\end{abstract}
\section{Introduction}
The following conjecture first appeared in \cite{Ful71} is known as
 the
Fulkerson Conjecture, see \cite{Sey}.
\begin{Conj}\label{Conjecture:Fulkerson}
If $G$ is a bridgeless cubic graph, then there exist $6$ perfect matchings $M_1,\ldots,M_6$ of $G$
with the property that every edge of $G$ is contained in exactly two of $M_1,\ldots,M_6$.
\end{Conj}

A consequence of the Fulkerson conjecture would be that
every bridgeless cubic graph has $3$ perfect matchings with empty
intersection (take $3$ distinct  of the $6$ perfect matchings given by the
conjecture). The following weakening of this (also suggested by
Berge) is still open.

\begin{Conj}\label{Conjecture:Berge2}
There exists a fixed integer $ k $ such that every
bridgeless cubic graph has a list of $ k $ perfect matchings with
empty intersection.
\end{Conj}

Fan and Raspaud \cite{FanRas} conjectured that any bridgeless cubic
graph can be provided with three perfect matchings with empty
intersection(we shall say also {\em non intersecting perfect
matchings}).

\begin{Conj}\cite{FanRas} \label{Conjecture:FanRaspaud} Every
bridgeless cubic graph contains perfect matching $M_1$, $M_2$, $M_3$
such that
$$M_1 \cap M_2 \cap M_3 = \emptyset$$
\end{Conj}

The following Conjecture is due to M\'a\u{c}ajov\'a and \u{S}koviera in \cite{MacSko07}
\begin{Conj}\label{Conj:MakajovaSkoviera}
Every bridgeless cubic graph has two perfect matchings $ M_1 $, $ M_2 $ such that $ M_1 \cap M_2 $ does not contain
an odd edge cut.
\end{Conj}

 A {\em join} in a graph $G$ is a set $J \subseteq
E(G)$ such that the degree of every vertex in G has the same parity
as its degree in the graph $(V (G), J)$. A perfect matching being a
particular join in a cubic graph Kaiser and Raspaud conjectured in
\cite{KaiRas}
\begin{Conj}\cite{KaiRas} \label{Conjecture:KaiserRaspaud} Every
bridgeless cubic graph admits two perfect matchings $M_1$, $M_2$ and
a join $J$ such that
$$M_1 \cap M_2 \cap J = \emptyset$$
\end{Conj}
As a matter of fact Conjectures \ref{Conj:MakajovaSkoviera} and \ref{Conjecture:KaiserRaspaud} are equivalent. Equivalence comes from the fact that a set of edges contains a join
if and only if this set intersects all odd edge cuts.

If true Conjecture \ref{Conjecture:Fulkerson} implies Conjecture
\ref{Conjecture:FanRaspaud} which itself implies Conjectures \ref{Conj:MakajovaSkoviera} and
\ref{Conjecture:KaiserRaspaud}. All those conjectures being obviously
true for cubic graphs with chromatic index $3$, it is useful  to
consider the following parameter for cubic graphs. The {\em oddness}
of a cubic graph $G$ is the minimum number of odd circuits in a
2-factor of $G$.  In \cite{KaiRas} Kaiser and Raspaud proved that
Conjecture \ref{Conjecture:KaiserRaspaud} holds true for bridgeless
cubic graph of oddness  two.

In this paper, we consider Conjecture \ref{Conj:MakajovaSkoviera}. We prove that a minimal
counterexample to Conjecture
\ref{Conjecture:KaiserRaspaud} must have at least $50$ vertices.

Moreover, we prove that Conjectures \ref{Conj:MakajovaSkoviera} and
\ref{Conjecture:KaiserRaspaud} hold true while the order of the
graph is less than a function of the cyclic edge connectivity.
Finally, we give a refining of Kaiser and Raspaud result
\cite{KaiRas} when considering cubic bridgeless traceable graphs.

When $A$ is a set of edges, $V(A)$ will denote the set of vertices that are an endpoint of some edge in $A$.
 If $M$ is a perfect matching of a cubic graph $G=(V,E)$, then denote by $G_M$ the $2$-factor $G_M=(V,E-M)$.
 When $X$ is a set of vertices, $\delta X$ denotes the set of edges with precisely one end in $X$.
 An edge cut is a set of edges whose removal renders the graph disconnected and which is inclusion-wise
 minimal with this property. The {\em cyclic edge connectivity} of a cubic graph is the size of a
smallest edge cut in a graph such that at least two of the connected
components contain cycles. A graph $G$ is said to be traceable
whenever $G$ has a {\em Hamiltonian path} that is a path which
visits each vertex exactly once.

{\bf Notations $\prec_W$ and $W(z,t)$, concatenation of sub-walks}.
Let $W$ be a walk of $G$. Writing $W=x\ldots y$ induces a
natural order on the vertices of $W$, let us denote $\prec_W$ this
order. When $W=x\ldots y$, $W$ will be said to {\em start} with $x$
and to {\em end} with $y$.When $z$ and $t$ are vertices of $W$ such
that $z\prec_W t$, the sub-walk $z\ldots t$ of $W$ whose endpoints
are $z$ and $t$ will be denoted $W(z,t)$. When a walk $W$
($W=W(x,y)$) and a walk $W'$ ($W'=W'(x',y')$) have a common vertex,
say $a$, we can {\em concatenate} the sub-walks $W(x,a)$ and
$W'(a,y')$ in order to obtain another walk say $W''$, also denoted
$W(x,a)+W'(a,y')$, such that $W''(x,a)=W(x,a)$ and
$W''(a,y')=W'(a,y')$.

 For basic graph-theoretic terms, we refer the
reader to Bondy and Murty \cite{BonMur08}.

\section{Preliminary results}\label{Sec:Préliminaires}
\subsection{Fractional perfect matchings}
The following result belongs to folklore
\begin{Thm}\label{Thm:3ColoriablePerfectMathingsEtOddCutSet}
Let $G$ be a cubic bridgeless graph.
$G$ is $3$-edge colourable if and only if there is a perfect matching in $G$ that does not contain any odd edge cut.
\end{Thm}
\begin{Prf}
Assume that $G$ has a $3$-edge colouring using colours $\alpha$, $\beta$ and $\gamma$. Let $M_{\alpha}$ be the set
of edges coloured with $\alpha$, if $M_{\alpha}$ contains an odd edge cut, there must be a partition $(V_1,V_2)$ of
$V(G)$ into two odd sets such that the edges of $X$ have one end in $V_1$ and the other end in $V_2$. Since all
the edges of $X$ are coloured with $\alpha$, the set of edges in $V_1$ coloured with $\beta$ must be a perfect
matching in $V_1$, a contradiction since $V_1$ has an odd number of vertices.

Conversely, consider a perfect matching $M$ that does not contain any odd edge cut. Suppose that the $2$-factor
$G_M$ contains an odd cycle $C$, thus $\delta C$ is an odd edge cut entirely contained in $M$, a contradiction.
Consequently $G_M$ does not contain any odd cycle, it follows that $G$ is $3$-edge colourable.
\end{Prf}

In order to prove Conjecture \ref{Conj:MakajovaSkoviera} for bridgeless cubic graphs with few vertices, we will consider the notion of
{\em fractional perfect matching} as used in \cite{KaiKraNor05}.

For a graph $G=(V,E)$, a vector $w$ of $\mathbb{R}^E$ is said to be a fractional perfect matching whenever
$w$ satisfies the following properties (the entry of $w$ corresponding to $e\in E$ being denoted $w(e)$ and
$w(A)=\Sigma_{e\in A} w(e)$, for $A\subseteq E$)~:
\newline $\bullet$ $0\leq w(e)\leq 1$ for each edge $e$ of $G$
\newline $\bullet$ $w(\delta\{v\})=1$ for each vertex $v$ of $G$
\newline $\bullet$ $w(\delta X)\geq 1$ for each $X\subseteq V$ of odd cardinality.

The perfect matching polytope is the convex hull of the set of
incidence vectors of perfect matchings of  $G$.
In \cite{Edm65} Edmonds showed that a vector $w\in \mathbb{R}^E$
belongs to the perfect matching polytope of  $G$
if and only if it is a fractional perfect matching

Moreover, when $\chi^A$ denotes the characteristic vector of the edge set $A$ we will use the following tool~:
\begin{Lem}\label{Lem:KaiSerKralNourine}\cite{KaiKraNor05}
If $w$ is a fractional perfect matching in a graph  $G = (V,E)$ and $c\in \mathbb{R}^E$, then
$G$ has a perfect matching $M$ such that $c.\chi^M\geq c.w$
where $.$ denotes the scalar product. Moreover, there exists such a
perfect matching $M$ that contains exactly one edge of each edge cut $C$
with $w(C) = 1$.
\end{Lem}

It is shown in \cite{KaiKraNor05}, among other results, that
 there must exist a perfect matching $M_1$ that intersects all
edge cuts of size $3$ into a single edge and a perfect matching
$M_2$ such that $|M_2-M_1|\geq \frac{4}{15}|E(G)|$. When the graph has $n$ vertices, since the size of a perfect matching is precisely $\frac{n}{2}$, it must be pointed out that $|M_1\cap M_2|\leq \frac{n}{10}$.

Observe that there is an alternate proof of Theorem
\ref{Thm:3ColoriablePerfectMathingsEtOddCutSet} in terms of
fractional perfect matchings  . Consider
indeed a perfect matching $M$ that does not contain any odd
edge cut. We define a fractional perfect matching as follows~:
$w(e)=0$ when $e\in M$ and $w(e)=\frac{1}{2}$ otherwise. Given an
odd set of vertices, say $X$, $\delta X$ is an odd edge cut which
intersects $M$ in a odd number of edges, since $\delta X\nsubseteq
M$, $w$ is a fractional perfect matching. By Lemma
\ref{Lem:KaiSerKralNourine} there is a perfect matching $M'$ such
that
$$c.\chi ^{M'}\geqslant c.w=\frac{1}{2}\times \frac{2}{3}\times |E|=\frac{n}{2}.$$
When $c=1-\chi^{M}$, since $c.\chi ^{M'}=|M'\backslash M|$ we have $|M'\backslash M|=\frac{n}{2}=|M'|$ and thus $M\cap M'=\emptyset$.
It follows that $\chi'(G)=3$.
\subsection{Balanced perfect matchings}
Let $M$ be a perfect matching of a  cubic graph and let $\mathcal
C=\{C_1,C_2 \ldots C_k\}$ be the 2-factor $G_M$. $A \subseteq M$ is
a {\em balanced} $M-$matching  whenever there is a perfect matching
$M'$ such that $M \cap M' =A$. That means that each odd cycle of
$\mathcal C$ is incident to an odd number of edges in $A$ and the
sub-paths determined by the ends of $A\cap M'$ on the cycles of $\mathcal C$ incident to $A$ have odd lengths.

\begin{Lem}\label{Lem:CouplageEquilibréG_M-V(A)}
Let $M$ be a perfect matching of a cubic graph $G$. A matching $A$ is a balanced $M$-matching if and only if the
connected components of $G_{M-A}$ are either odd paths or even cycles.
\end{Lem}
\begin{Prf}
Since $G_M$ is a $2$-factor of $G$, the connected components of the subgraph induced by $V(G)-V(A)$ must be cycles or paths.
Since $A$ is a  balanced $M$-matching, the connected components
of this graph must be even cycles or odd paths.

Conversely, assume that the connected components of $G_M-V(A)$ are odd paths or even cycles. Let $A'$ be a perfect matching of $G_M-V(A)$, we set $M'=A\cup A'$ and we are done.
\end{Prf}

\begin{Lem}\label{Lemma:FondamentalDisjointsBalanced}
A bridgeless cubic graph contains $3$ non intersecting perfect
matching if and only if there is a perfect matching $M$ and two
balanced disjoint balanced $M-$matchings.
\end{Lem}
\begin{Prf} Assume that $M_1$, $M_2$, $M_3$ are three perfect
matchings of $G$ such that $M_1 \cap M_2 \cap M_3 = \emptyset$. Let
$M=M_1$, $A=M_1 \cap M_2$ and $B=M_1 \cap M_3$. Since $A\cap B=M_1
\cap M_2 \cap M_3$, $A$ and $B$ are two balanced $M-$matchings with
empty intersection.

Conversely, assume that $M$ is a perfect matching and that $A$ and
$B$ are two balanced $M-$matchings with empty intersection. Let
$M_1=M$, $M_2$ be a perfect matching such that $M_2 \cap M_1=A$
 and $M_3$ be a perfect
matching such that $M_3 \cap M_1=B$. We have $M_1 \cap M_2 \cap
M_3=A \cap B$ and the three perfect matchings $M_1$, $M_2$ and $M_3$
have an empty intersection.
\end{Prf}

\section{On cubic graphs with few vertices}
We first prove that Conjecture \ref{Conj:MakajovaSkoviera} holds true for bridgeless cubic graphs having less
than $50$ vertices
\begin{Thm}\label{Thm:ConjMSPourMoinsDe50}
Let $G$ be a bridgeless cubic graph of order $n<50$. There
 are perfect matchings $M$ and $M'$  such that
$M\cap M'$ does not contain any edge cut.
\end{Thm}
\begin{Prf}
We know from \cite{KaiKraNor05} that  there must exist a perfect
matching $M$ which intersects all edge cuts of size $3$ into a
single edge and a perfect matching $M'$ such that $|M\cap M'|\leq
\frac{n}{10}$. It is assumed $n<50$, thus $|M\cap M'|<5$.
Hence  an odd edge cut, say $C$  in
$M\cap M'$ must be of size $3$, but a such edge cut cannot exist
since  $M$ intersects $C$ in precisely one edge.
\end{Prf}

Let us now consider cyclic edge connectivity in cubic graphs.
\begin{Thm}\label{Thm:Cycliquement-k-ArêteConnexes}
Let $G$ be a cubic graph of order $n$ with cyclic edge connectivity
 $ k \geq 3$. One of the following holds.
\begin{enumerate}
\item There are two perfect matchings $M$ and $M'$ such that
$|M\cap M'|\leq \frac{n}{2(2\lfloor\frac{k}{2}\rfloor+3)}$.
\item For all perfect matching $M$ there is an edge cut of size $2\lfloor\frac{k}{2}\rfloor+1$ entirely contained
in $M$.
\end{enumerate}
\end{Thm}
\begin{Prf}
For convenience we set $s=2\lfloor\frac{k}{2}\rfloor+3$. Let $M$
be a perfect matching that does not contain any odd edge
cut of size $s-2$. The graph being cyclically $k$-edge connected
$s-2$ is the minimum size of an odd edge cut in $G$. We set
$w(e)=\frac{1}{s}$ when $e\in M$ and $w(e)=\frac{s-1}{2s}$
otherwise. If $X$ is an odd set of vertices, $\delta X$ is an odd
edge cut of size at least $s-2$.
If $|\delta X|\geq s$ then $w(\delta X)\geq 1$. If $|\delta
X|=s-2$ then  there are at least $2$ edges of $\delta X$ which are
not in $M$ and $w(\delta X)\geq 1$ again.  Hence $w$ is a fractional
perfect matching.

Applying Lemma \ref{Lem:KaiSerKralNourine} with $c=1-\chi^M$
we get a perfect matching, say $M'$ such that
$c.\chi^{M'}\geq c.w=n\times \frac{s-1}{2s}$. Since
$c.\chi^{M'}=|M'-M|$ and $|M'|=\frac{n}{2}$ it follows that $|M\cap
M'|\leq \frac{n}{2s}$, as claimed.
\end{Prf}
\begin{Thm}\label{Thm:MakajovaSkovieraOrdrePeuDeSommets}
Let $G$ be a cubic graph of order $n$ with cyclic edge connectivity $k\geq 4$.
If $n<2(2\lfloor\frac{k}{2}\rfloor+3)(2\lfloor\frac{k}{2}\rfloor+1)$ then there are two perfect matchings whose
intersection does not contain any odd edge cut.
\end{Thm}
\begin{Prf}
Once again we denote $s=2\lfloor\frac{k}{2}\rfloor+3$. We can assume that every perfect matching contains an odd
edge cut of size $s-2$. Otherwise, from Theorem
\ref{Thm:Cycliquement-k-ArêteConnexes} there are two perfect
matchings whose intersection contains less than $\frac{n}{2s}=s-2$
edges and we are done.

Let $M$ be a perfect matching of $G$. We set $w(e)=\frac{1}{s-2}$
when $e\in M$ and $w(e)=\frac{s-3}{2(s-2)}$ otherwise. The weight of
an edge being at least $\frac{1}{s-2}$  and an
odd edge cut having at least $s-2$ edges, $w$ is a
fractional perfect matching. If $c=1-\chi^M$, by Lemma
\ref{Lem:KaiSerKralNourine} there is a perfect matching $M'$ which
intersects in a single edge every edge cut $C$ such that $w(C)=1$.

In addition we know that $c.\chi^{M'}\geq c.w$, in other words
$|M\cup M'|\geq \frac{2}{3}\times |E| \times
\frac{s-3}{2(s-2)}=\frac{n}{2}\times \frac{s-3}{s-2}$. Consequently $|M\cap M'|\leq\frac{n}{2(s-2)}$. Since $n<2s(s-2)$ we have that $|M\cap M'|\leq s$.

Assume that $M\cap M'$ contains an odd edge cut $C$. By the above
relation $|C|=s-2$ and then $w(C)=1$, a
contradiction since $M'$ intersects the edge cuts of size
$s-2$ in a single edge.
\end{Prf}

An example of consequence of Theorem \ref{Thm:MakajovaSkovieraOrdrePeuDeSommets} is that
Conjecture \ref{Conj:MakajovaSkoviera} and therefore Conjecture \ref{Conjecture:KaiserRaspaud} hold
true for cyclically $4$-edge-connected graphs having less than $70$ vertices.

\begin{Rem}
Kaiser, Kr\'{a}l and Norine in \cite{KaiKraNor05} showed that every bridgeless cubic graph contains two perfect matchings whose
intersection has at most $\frac{n}{10}$ edges. This result strengthen Fulkerson conjecture. Indeed, if we have a set of $6$ perfect matchings such that any edge of a bridgeless cubic graph is covered exactly twice by this set, we certainly have two of them whose intersection has at most $\frac{n}{10}$ edges. Observe that this upper bound would be implied by Conjecture \ref{Conjecture:Fulkerson}. A challenging question is thus to characterize the bridgeless cubic graphs for which $\frac{n}{10}$ is optimal. The Petersen graph is obviously such a graph, but no other graph is known with that property and we can conjecture that this is the only graph.

 Theorem \ref{Thm:Cycliquement-k-ArêteConnexes} above says
that in  a cyclically $4-$edge connected cubic graph, with
chromatic index $4$, either we can find two perfect matchings whose
intersection has at most $\frac{n}{14}$ edges or every two perfect
matchings has an intersection containing an odd cutset of size $5$,
a support to the above conjecture.
\end{Rem}

\section{On cubic traceable graphs}
In \cite{KaiRas} Kaiser and Raspaud proved that
Conjecture \ref{Conjecture:KaiserRaspaud} holds true for bridgeless cubic graph of oddness  two. In the following we prove a stronger result for cubic bridgeless traceable graphs.

\subsection{An auxiliary graph \label{SubSection:AuxiliaryGraph}}

Let us consider a Hamiltonian path of $G$. It will be convenient to denote the  vertices of $G$ as integers from $1$ to $n$ and the Hamiltonian path will be merely denoted $1 \ldots n$. Hence $ij$ ($ i\not = j \in \{1 \ldots n\})$ denotes an edge of $G$ while the edge joining $i$ to $i+1$($1 \leq i \leq n-1$) will be denoted $e_{i}$.

 Suppose that chromatic index of $G$ is $4$, we can colour the edges of $G$ in the following way. The edges $e_i$ ($1 \leq i \leq n-1$) of the Hamiltonian path are alternately coloured with $\alpha$ and $\beta$ (the first edge $e_1$ being coloured with $\alpha$). The remaining edges are coloured with $\gamma$ with the exception of one edge incident with $1$ and one edge incident with $n$.
These two edges are coloured by $\delta$. The set $M_{\alpha}$ of edges coloured with $\alpha$ is a perfect matching and the $2-$factor $G_{M_{\alpha}}=\{C_{1} \ldots C_{k}\}$ is composed of a set of even cycles whose edges are coloured $\beta$ or $\gamma$ and two odd cycles $C_{1}$ and $C_{k}$. Without loss of generality we suppose that $1$ is a vertex of $C_{1}$ and $n$ is a vertex of $C_{k}$.

\paragraph*{The edges $e_{min(C)}$ and $e_{max(C)}$.} For $C\in \{C_1,C_2\ldots C_k\}$ we denote $max(C)$
the greatest index $i$ such that $e_i$ is an edge of $C$, similarly $min(C)$ denotes the smallest index $i$
such that $e_i$ belongs to $C$. Observe that $max(C)$ and $min(C)$ are even numbers and that the corresponding
edges are coloured with $\beta$. Moreover the endpoints of $e_{min(C)}$ are $min(C)$ and $min(C)+1$ as well as
the endpoints of $e_{max(C)}$ are $max(C)$ and $max(C)+1$

Observe that $min(C)$ and  $max(C)$ are always defined and that
$min(C)=max(C)$ if and only if $C$ is a triangle.

\paragraph*{The sequence $(\Gamma_j)_{j=1\ldots h}$.} We define a sequence $(\Gamma_j)_{j=1\ldots h}$,
$2\leq h\leq k$ of members of $\{C_1,\ldots C_{k}\}$ as follows~:

\begin{itemize}
  \item  We set $\Gamma_1=C_1$.
  \item  If $max(\Gamma_j)< min(C_k)$, since the edge $e_{max(\Gamma_j)+1}$ is not a bridge there
is a cycle $C$ in $G_{M_{\alpha}}$ with $min(C)< max(\Gamma_j)<
max(C)$. Among all such cycles, let us denote by $\Gamma_{j+1}$ the
cycle $C$ for which $max(C)$ is maximum.
  \item If
$max(\Gamma_j)> min(C_k)$, we set $h=j+1$ and $\Gamma_h=C_k$.
\end{itemize}

Observe that by construction, when $h=2$, we have
$$1<min(\Gamma_{2})=min(C_k)<max(C_1)=max(\Gamma_1)<n$$  and when $h>2$, we have $$min(\Gamma_j)< max(\Gamma_{j-1})< min(\Gamma_{j+1})<
max(\Gamma_{j}) \qquad 1<j<h$$ and
$$min(\Gamma_h)=min(C_k)<n$$.

\begin{figure}
\begin{center}
"
\begin{picture}(0,0)%
\includegraphics[scale=0.5]{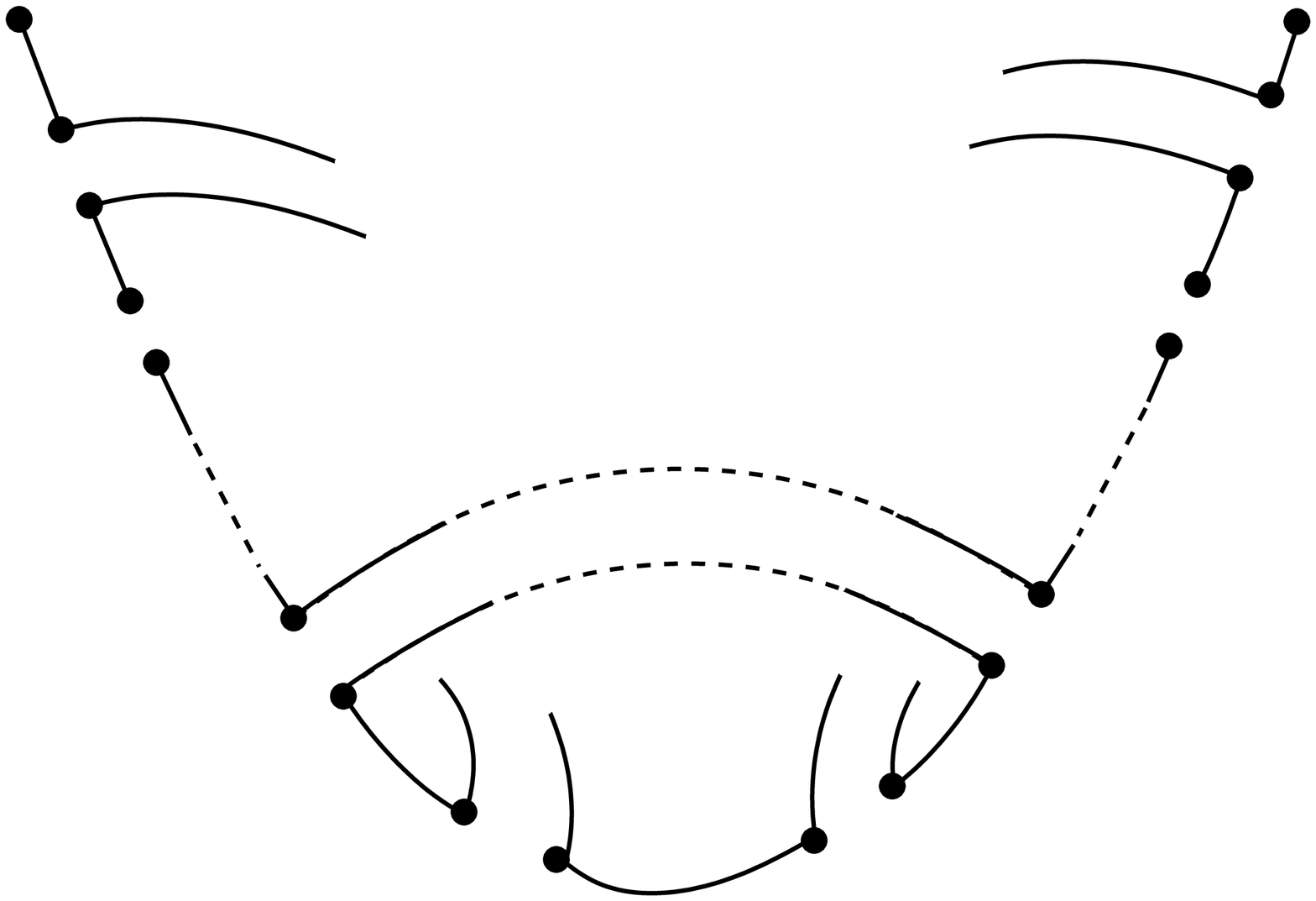}%
\end{picture}%
\setlength{\unitlength}{1973sp}%
\begingroup\makeatletter\ifx\SetFigFont\undefined%
\gdef\SetFigFont#1#2#3#4#5{%
  \reset@font\fontsize{#1}{#2pt}%
  \fontfamily{#3}\fontseries{#4}\fontshape{#5}%
  \selectfont}%
\fi\endgroup%
\begin{picture}(9566,6774)(-343,-6946)
\put(-328,-615){\makebox(0,0)[rb]{\smash{{\SetFigFont{9}{10.8}{\rmdefault}{\mddefault}{\updefault}{\color[rgb]{0,0,0}$1$}%
}}}}
\put(9208,-400){\makebox(0,0)[lb]{\smash{{\SetFigFont{9}{10.8}{\rmdefault}{\mddefault}{\updefault}{\color[rgb]{0,0,0}$n$}%
}}}}
\put(-12,-1347){\makebox(0,0)[rb]{\smash{{\SetFigFont{9}{10.8}{\rmdefault}{\mddefault}{\updefault}{\color[rgb]{0,0,0}$min(\Gamma_2)$}%
}}}}
\put(145,-1935){\makebox(0,0)[rb]{\smash{{\SetFigFont{9}{10.8}{\rmdefault}{\mddefault}{\updefault}{\color[rgb]{0,0,0}$min(\Gamma_2)+1$}%
}}}}
\put(1651,-4760){\makebox(0,0)[rb]{\smash{{\SetFigFont{9}{10.8}{\rmdefault}{\mddefault}{\updefault}{\color[rgb]{0,0,0}$min(\Gamma_j)$}%
}}}}
\put(1952,-5334){\makebox(0,0)[rb]{\smash{{\SetFigFont{9}{10.8}{\rmdefault}{\mddefault}{\updefault}{\color[rgb]{0,0,0}$min(\Gamma_j)+1$}%
}}}}
\put(8321,-3025){\makebox(0,0)[lb]{\smash{{\SetFigFont{9}{10.8}{\rmdefault}{\mddefault}{\updefault}{\color[rgb]{0,0,0}$min(\Gamma_h)$}%
}}}}
\put(5524,-6725){\makebox(0,0)[lb]{\smash{{\SetFigFont{9}{10.8}{\rmdefault}{\mddefault}{\updefault}{\color[rgb]{0,0,0}$min(\Gamma_{j+1})$}%
}}}}
\put(6455,-6022){\makebox(0,0)[lb]{\smash{{\SetFigFont{9}{10.8}{\rmdefault}{\mddefault}{\updefault}{\color[rgb]{0,0,0}$min(\Gamma_{j+1})+1$}%
}}}}
\put(518,-2537){\makebox(0,0)[rb]{\smash{{\SetFigFont{9}{10.8}{\rmdefault}{\mddefault}{\updefault}{\color[rgb]{0,0,0}$max(\Gamma_1)$}%
}}}}
\put(661,-3039){\makebox(0,0)[rb]{\smash{{\SetFigFont{9}{10.8}{\rmdefault}{\mddefault}{\updefault}{\color[rgb]{0,0,0}$max(\Gamma_1)+1$}%
}}}}
\put(7058,-5248){\makebox(0,0)[lb]{\smash{{\SetFigFont{9}{10.8}{\rmdefault}{\mddefault}{\updefault}{\color[rgb]{0,0,0}$max(\Gamma_j)$}%
}}}}
\put(2913,-6281){\makebox(0,0)[rb]{\smash{{\SetFigFont{9}{10.8}{\rmdefault}{\mddefault}{\updefault}{\color[rgb]{0,0,0}$max(\Gamma_{j-1})$}%
}}}}
\put(3845,-6826){\makebox(0,0)[rb]{\smash{{\SetFigFont{9}{10.8}{\rmdefault}{\mddefault}{\updefault}{\color[rgb]{0,0,0}$max(\Gamma_{j-1})+1$}%
}}}}
\put(8822,-1691){\makebox(0,0)[lb]{\smash{{\SetFigFont{9}{10.8}{\rmdefault}{\mddefault}{\updefault}{\color[rgb]{0,0,0}$max(\Gamma_{h-1})$}%
}}}}
\put(8979,-1146){\makebox(0,0)[lb]{\smash{{\SetFigFont{9}{10.8}{\rmdefault}{\mddefault}{\updefault}{\color[rgb]{0,0,0}$max(\Gamma_{h-1})+1$}%
}}}}
\put(8549,-2436){\makebox(0,0)[lb]{\smash{{\SetFigFont{9}{10.8}{\rmdefault}{\mddefault}{\updefault}{\color[rgb]{0,0,0}$min(\Gamma_h)+1$}%
}}}}
\put(7459,-4602){\makebox(0,0)[lb]{\smash{{\SetFigFont{9}{10.8}{\rmdefault}{\mddefault}{\updefault}{\color[rgb]{0,0,0}$max(\Gamma_j)+1$}%
}}}}
\end{picture}%

\end{center}
\caption{An auxiliary graph (with $h>2$)}
\label{Fig:GrapheAuxiliaire}
\end{figure}

\paragraph*{An auxiliary graph $H$.} We consider an auxiliary graph $H$, where $V(H)=V(G)$
and $E(H)$ is obtained from $E(G)$ as follows (see Figure
\ref{Fig:GrapheAuxiliaire})~:

\begin{itemize}
  \item We delete the edges $e_{max(\Gamma_1)}$ and $e_{min(\Gamma_h)}$ and the edges
$e_{min(\Gamma_j)}$ and $e_{max(\Gamma_j)}$ of each cycle $\Gamma_j$
($1<j<h$) .
  \item Since $\Gamma_j$ ($1<j<h$) is an even cycle, when deleting the edges $e_{min(\Gamma_j)}$ and
$e_{max(\Gamma_j)}$, we get two odd paths with one end in
$\{min(\Gamma_j), min(\Gamma_j)+1\}$ and the other end in
$\{max(\Gamma_j), max(\Gamma_j)+1\}$, namely $P_j$ and $P'_j$. We
put in $E(H)$ two new edges (denoted in the following  as {\em
additional} edges) one edge connecting the endpoints of $P_j$ while
the endpoints of the other edge are the endpoints of $P'_j$. We will
say in the following that the first edge {\em represents} the path
$P_j$ while the other one {\em represents} the path $P'_j$
\item Finally, we delete all the edges of $G$ being coloured with $\gamma$ and $\delta$ (that is the chords of the Hamiltonian path).
\end{itemize}

All the vertices of $H$ have degree $2$ except $6$ vertices, namely
$1$, $max(\Gamma_1)$, $max(\Gamma_1)$+1, $min(\Gamma_h)$,
$min(\Gamma_h)+1$, $n$ which have degree $1$. Thus the connected
components of $H$ are precisely $3$ paths whose endpoints are
members of $\{1, max(\Gamma_1), max(\Gamma_1)+1, min(\Gamma_h),
min(\Gamma_h)+1, n\}$

{\bf Notation.}

Two walks of $G$ will be said $\alpha$-disjoint whenever those walks do not share any edge coloured with $\alpha$.

In the following, when a cubic brigeless traceable $G$ graph is given we will consider
the graph $G$ together with the edge-colouring defined above, the sequence $(\Gamma_j)_{j=1\ldots h}$, the auxiliary graph $H$ and all related notations.

\subsection{Technical lemmas}

\begin{Lem}\label{Lem:TroisWalksDeC1ACn}
Let $G$ be a cubic bridgeless traceable graph such that
$\chi'(G)=4$. There are in $G$ three pairwise $\alpha$-disjoint odd
walks, $W_1$, $W_2$, $W_3$ such that, for $i\in \{1,2,3\}$~:

\begin{itemize}
  \item $W_i$ has one endpoint say $q_i$ in $C_1$ and the other endpoint $q'_i$ in $C_k$
  \item $W_i$ does not share any edge with $C_1$ nor $C_k$
\end{itemize}
\end{Lem}
\begin{Prf}
Those walks will be derived from the connected components of the
auxiliary graph $H$. As a matter of fact, when $h=2$, $H$ is reduced
to $3$ sub-paths of $P$, namely: $Q_{1}$:
$1...min(\Gamma_{2})$,$Q_{2}$: $min(\Gamma_{2})+1....
max(\Gamma_{1})$ and $Q_{3}$:$max(\Gamma_{1})+1....n$. We can thus
suppose that $h>2$.

Let $Q=x_1\ldots x_r$ be a connected component of $H$ with its
endpoint $x_1$ in $\{1, max(\Gamma_1), max(\Gamma_1)+1\}$, we will
prove that the other endpoint $x_r$ of $Q$ belongs to
$\{min(\Gamma_h), min(\Gamma_h)+1, n\}$. Let $x$ be the maximum
index in $\{1,\ldots n\}$ of a vertex of $Q$.
\begin{Clm}\label{Claim:xApresGammaj}
$x> max(\Gamma_{h-2})$.
\end{Clm}
\begin{PrfClaim}
It is easy to check that $x> max(\Gamma_1)$.

If $x>max(\Gamma_{j-1})$ and $x\leq max(\Gamma_j)$ for $1< j < h-1$, then the vertex $x$ must be a vertex of one of the $2$ sub-paths $max(\Gamma_{j-1})+1\ldots min(\Gamma_{j+1})$ or
$min(\Gamma_{j+1})+1\ldots max(\Gamma_j)$ of $P$, thus $x=min(\Gamma_{j+1})$ or $x=max(\Gamma_j)$. In both cases, since $j<h-1$ there must be in $Q$ one vertex of $\{max(\Gamma_{j+1}),max(\Gamma_{j+1})+1\}$. But those vertices have
an index greater than $x$, a contradiction. Thus $x> max(\Gamma_{h-2})$.
\end{PrfClaim}

\begin{Clm}\label{Claim:TroisCheminsMixtes}
The connected components of $H$ are odd paths with one end in \\ $\{1, max(\Gamma_1), max(\Gamma_1)+1\}$ and
the other end in $\{min(\Gamma_h), min(\Gamma_h)+1, n\}$.
\end{Clm}
\begin{PrfClaim}
From Claim \ref{Claim:xApresGammaj}, $x> max(\Gamma_{h-2})$ and
either $x=min(\Gamma_h)=x_r$ or $x=max(\Gamma_{h-1})$, in that case
$x_r=min(\Gamma_h)+1$, or $x=n=x_r$. Consequently no connected
component of $H$ can be a path with both ends in $\{1,
max(\Gamma_1), max(\Gamma_1)+1\}$, the Claim follows.
\end{PrfClaim}

To a path, say $Q_s$ ($s\in\{1,2,3\}$), of $H$ we can associate an
odd walk $W_s$ of $G$ as follows~:

\begin{itemize}
  \item Let $q_s$ be the last vertex of $Q_s$ that belongs to $C_1$ when
running on $Q_s$ from its endpoint in $\{1, max(C_1),
max(C_1)+1\}$. Similarly let $q'_s$ be the last vertex of $Q_s$ that
belongs to $C_k$ when running on $Q_s$ from its endpoint in $\{n,
min(C_k), min(C_k)+1\}$. Let $Q'_s$ be the sub-path of $Q_s$ whose
endpoints are $q_s$ and $q'_s$.
  \item Each  {\em additional} edge of $Q'_s$  represents some odd sub-path ($P_j$ or $P'_j$) of some
cycle $\Gamma_j$. The walk $W_s$ is obtained from the path $Q'_s$ by replacement of each additional edge with the sub-path that it represents.
\end{itemize}
The walks $W_s$ ($s\in\{1,2,3\}$) defined above are
$\alpha$-disjoint walks. Moreover those walks have one end
which belongs to $C_1$ and the other end which belongs to $C_k$,
have no edge of $C_1$ nor of $C_k$ while their end-edges are
coloured with $\alpha$.
\end{Prf}

\begin{figure}
\begin{center}
\begin{picture}(0,0)%
\special{psfile=QetQPrime.ps}%
\end{picture}%
\setlength{\unitlength}{2072sp}%
\begingroup\makeatletter\ifx\SetFigFont\undefined
\def\x#1#2#3#4#5#6#7\relax{\def\x{#1#2#3#4#5#6}}%
\expandafter\x\fmtname xxxxxx\relax \def\y{splain}%
\ifx\x\y   
\gdef\SetFigFont#1#2#3{%
  \ifnum #1<17\tiny\else \ifnum #1<20\small\else
  \ifnum #1<24\normalsize\else \ifnum #1<29\large\else
  \ifnum #1<34\Large\else \ifnum #1<41\LARGE\else
     \huge\fi\fi\fi\fi\fi\fi
  \csname #3\endcsname}%
\else
\gdef\SetFigFont#1#2#3{\begingroup
  \count@#1\relax \ifnum 25<\count@\count@25\fi
  \def\x{\endgroup\@setsize\SetFigFont{#2pt}}%
  \expandafter\x
    \csname \romannumeral\the\count@ pt\expandafter\endcsname
    \csname @\romannumeral\the\count@ pt\endcsname
  \csname #3\endcsname}%
\fi
\fi\endgroup
\begin{picture}(11346,4557)(-229,-5290)
\put(598,-4411){\makebox(0,0)[lb]{\smash{\SetFigFont{6}{7.2}{rm}{}{}$Q$}}}
\put(9534,-3204){\makebox(0,0)[lb]{\smash{\SetFigFont{6}{7.2}{rm}{}{}$e_{max(\Gamma_j)}$}}}
\put( 50,-3088){\makebox(0,0)[lb]{\smash{\SetFigFont{6}{7.2}{rm}{}{}$e_{min(\Gamma_j)}$}}}
\put(8722,-2215){\makebox(0,0)[lb]{\smash{\SetFigFont{6}{7.2}{rm}{}{}$\gamma$}}}
\put(8889,-4282){\makebox(0,0)[lb]{\smash{\SetFigFont{6}{7.2}{rm}{}{}$\gamma$}}}
\put(-39,-893){\makebox(0,0)[lb]{\smash{\SetFigFont{6}{7.2}{rm}{}{}$x'$}}}
\put(10928,-1724){\makebox(0,0)[lb]{\smash{\SetFigFont{6}{7.2}{rm}{}{}$y'$}}}
\put(-136,-5230){\makebox(0,0)[lb]{\smash{\SetFigFont{6}{7.2}{rm}{}{}$x$}}}
\put(11117,-4572){\makebox(0,0)[lb]{\smash{\SetFigFont{6}{7.2}{rm}{}{}$y$}}}
\put(4852,-3117){\makebox(0,0)[lb]{\smash{\SetFigFont{12}{14.4}{rm}{}{}$\Gamma_j$}}}
\put(510,-1518){\makebox(0,0)[lb]{\smash{\SetFigFont{6}{7.2}{rm}{}{}$Q'$}}}
\put(1964,-2285){\makebox(0,0)[lb]{\smash{\SetFigFont{6}{7.2}{rm}{}{}$\gamma$}}}
\put(1787,-4146){\makebox(0,0)[lb]{\smash{\SetFigFont{6}{7.2}{rm}{}{}$\gamma$}}}
\end{picture}
\caption{The walks $Q$ and $Q'$ that intersect $\Gamma_j$}\label{fig:QetQ'}
\end{center}
\end{figure}
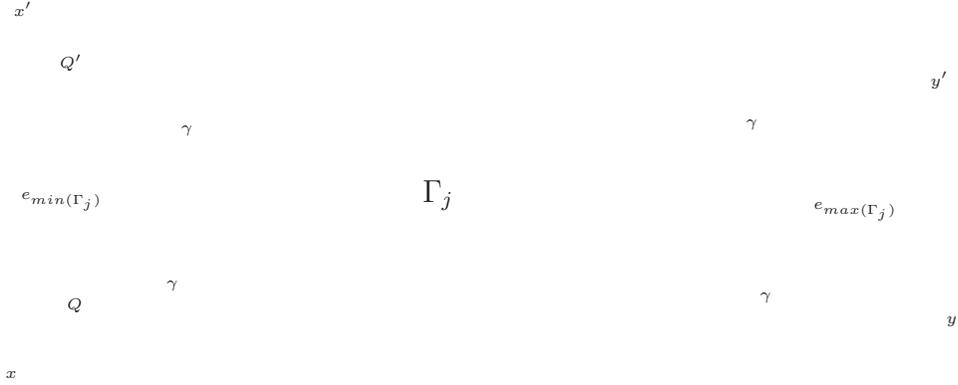

We shall deal  in the next subsection  with the particular
case where the sequence $(\Gamma_j)_{j=1\ldots h}$ contains only the
two odd cycles $C_{1}$ and $C_{k}$, see Proposition
\ref{prop:FanRaspaudQuandOna2Cycles}. Hence, we assume in
the sequel of this subsection that $h>2$, we give below some notations in order to
describe the construction from $W_1$, $W_2$ and $W_3$ of
new walks which intersect the even cycles of the sequence.

We intend to derive $Q$ into a walk which set of $\alpha$-edges can
be extended into a perfect matching on $\Gamma_j$.

An odd subpath of $C_{i}$ whose end edges are coloured with $\gamma$ is a {\em $\gamma$-chain} and we define analogously a {\em $\beta$-chain}. A walk $W$ will be
said to {\em well-intersect} a cycle $C$ of $G_{M_{\alpha}}$ when either $W\cap C=\emptyset$ or the set of endpoints of the
$\alpha$-edges of $W$ which belong to $C$, say $\{a_1\ldots a_p\}$ in that order around $C$, are such that the consecutive subpaths $\{[a_i\ldots a_{i+1}]\}_{0<i<p}$ are odd..

\begin{Lem} \label{Lemma:WellIntersectedNotInvolved}
Let $W_1$, $W_2$, $W_3$ be the three walks described in
Lemma \ref{Lem:TroisWalksDeC1ACn} then, each even cycle of
$G_{M_{\alpha}}$  not involved in the sequence
$(\Gamma_j)_{j=1\ldots h}$ is well-intersected  by $W_{i}$ ($i \in
\{1,2,3\}$).
\end{Lem}

\begin{Prf} Let $C$ be an even cycle not involved in the sequence
$(\Gamma_j)_{j=1\ldots h}$. The walks $W_{i}$ $i \in \{1,2,3\}$
possibly intersect this cycle in using only edges coloured
$\beta$. The $\alpha$ edges of $W_{i}$ with an end in $C$ determine
thus a set of $\gamma$-chains or $\beta$-chains. The result follows.
\end{Prf}

\begin{Lem}\label{Lemma:WellIntersectedGammajNotInvolved}
Let $W_1$, $W_2$, $W_3$ be the three walks described in
Lemma \ref{Lem:TroisWalksDeC1ACn} then, for each even cycle of the
sequence $(\Gamma_j)_{j=1\ldots h}$ the walk which do not use the
vertices $max(\Gamma_{j})$ and $max(\Gamma_{j})+1$ well-intersects
$\Gamma_{j}$.
\end{Lem}

\begin{Prf} Assume without loss of generality that  $W_1$ does not use neither $max(\Gamma_{j})$ nor
$max(\Gamma_{j})+1$, then, by construction, $W_1$ has not been
obtained by replacement of the additional edges representing the two
paths $P_{i}$ or $P'_{i}$ of $C_i=\Gamma_j$. Hence $W_{1}$ possibly intersects
$\Gamma_{j}$ by using only edges coloured $\beta$. The $\alpha$
edges of $W_{1}$ with an end in $\Gamma_{j}$ determine thus a set
of $\gamma$-chains or $\beta$-chains. The result follows.
\end{Prf}

Given an even cycle of the sequence
$(\Gamma_j)_{j=1\ldots h}$ say $\Gamma_j$, there are precisely two
walks in $\{W_1, W_2, W_3\}$ say $Q=Q(x,y)$ and $Q'(x',y')$,
$x\in\Gamma_1$, $y \in \Gamma_h$, $x'\in \Gamma_1$, $y' \in
\Gamma_h$, both of them containing a subpath of $\Gamma_j$ with
end-edges coloured with $\gamma$ (see Figure \ref{fig:QetQ'}). Morover both $Q$ and $Q'$ contain precisely one vertex of $\{max(\Gamma_j),max(\Gamma_j)+1\}$.

The first vertex of $Q$ (resp. $Q'$) following the order given by
$\prec_Q$ (resp. $\prec_{Q'})$ that belongs to $\Gamma_j$ is denoted
$x_j$ (resp. $x'_j$).

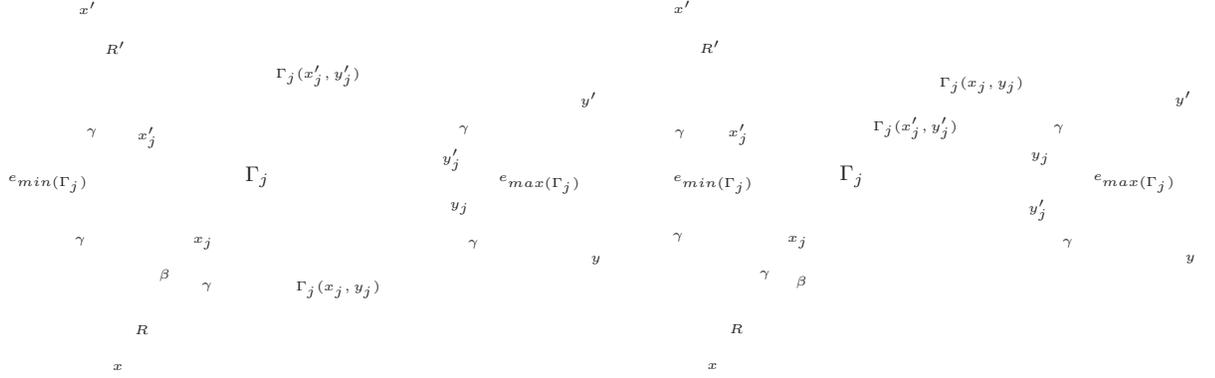
\begin{figure}
\begin{center}
\begin{picture}(0,0)%
\special{psfile=RetRPrime.ps}%
\end{picture}%
\setlength{\unitlength}{1367sp}%
\begingroup\makeatletter\ifx\SetFigFont\undefined
\def\x#1#2#3#4#5#6#7\relax{\def\x{#1#2#3#4#5#6}}%
\expandafter\x\fmtname xxxxxx\relax \def\y{splain}%
\ifx\x\y   
\gdef\SetFigFont#1#2#3{%
  \ifnum #1<17\tiny\else \ifnum #1<20\small\else
  \ifnum #1<24\normalsize\else \ifnum #1<29\large\else
  \ifnum #1<34\Large\else \ifnum #1<41\LARGE\else
     \huge\fi\fi\fi\fi\fi\fi
  \csname #3\endcsname}%
\else
\gdef\SetFigFont#1#2#3{\begingroup
  \count@#1\relax \ifnum 25<\count@\count@25\fi
  \def\x{\endgroup\@setsize\SetFigFont{#2pt}}%
  \expandafter\x
    \csname \romannumeral\the\count@ pt\expandafter\endcsname
    \csname @\romannumeral\the\count@ pt\endcsname
  \csname #3\endcsname}%
\fi
\fi\endgroup
\begin{picture}(21350,6695)(550,-6605)
\put(17441,-1404){\makebox(0,0)[lb]{\smash{\SetFigFont{5}{6.0}{rm}{}{}$\Gamma_j(x_j,y_j)$}}}
\put(20234,-3104){\makebox(0,0)[lb]{\smash{\SetFigFont{5}{6.0}{rm}{}{}$e_{max(\Gamma_j)}$}}}
\put(12617,-3121){\makebox(0,0)[lb]{\smash{\SetFigFont{5}{6.0}{rm}{}{}$e_{min(\Gamma_j)}$}}}
\put(9451,-3121){\makebox(0,0)[lb]{\smash{\SetFigFont{5}{6.0}{rm}{}{}$e_{max(\Gamma_j)}$}}}
\put(550,-3105){\makebox(0,0)[lb]{\smash{\SetFigFont{5}{6.0}{rm}{}{}$e_{min(\Gamma_j)}$}}}
\put(2894,-2356){\makebox(0,0)[lb]{\smash{\SetFigFont{5}{6.0}{rm}{}{}$x'_{j}$}}}
\put(3906,-4232){\makebox(0,0)[lb]{\smash{\SetFigFont{5}{6.0}{rm}{}{}$x_{j}$}}}
\put(4852,-3117){\makebox(0,0)[lb]{\smash{\SetFigFont{8}{9.6}{rm}{}{}$\Gamma_j$}}}
\put(8565,-3601){\makebox(0,0)[lb]{\smash{\SetFigFont{5}{6.0}{rm}{}{}$y_{j}$}}}
\put(8421,-2764){\makebox(0,0)[lb]{\smash{\SetFigFont{5}{6.0}{rm}{}{}$y'_{j}$}}}
\put(11117,-4572){\makebox(0,0)[lb]{\smash{\SetFigFont{5}{6.0}{rm}{}{}$y$}}}
\put(2447,-6545){\makebox(0,0)[lb]{\smash{\SetFigFont{5}{6.0}{rm}{}{}$x$}}}
\put(10928,-1724){\makebox(0,0)[lb]{\smash{\SetFigFont{5}{6.0}{rm}{}{}$y'$}}}
\put(1835,-87){\makebox(0,0)[lb]{\smash{\SetFigFont{5}{6.0}{rm}{}{}$x'$}}}
\put(4056,-5065){\makebox(0,0)[lb]{\smash{\SetFigFont{5}{6.0}{rm}{}{}$\gamma$}}}
\put(8889,-4282){\makebox(0,0)[lb]{\smash{\SetFigFont{5}{6.0}{rm}{}{}$\gamma$}}}
\put(8722,-2215){\makebox(0,0)[lb]{\smash{\SetFigFont{5}{6.0}{rm}{}{}$\gamma$}}}
\put(3289,-4882){\makebox(0,0)[lb]{\smash{\SetFigFont{5}{6.0}{rm}{}{}$\beta$}}}
\put(2309,-805){\makebox(0,0)[lb]{\smash{\SetFigFont{5}{6.0}{rm}{}{}$R'$}}}
\put(2852,-5886){\makebox(0,0)[lb]{\smash{\SetFigFont{5}{6.0}{rm}{}{}$R$}}}
\put(1756,-4231){\makebox(0,0)[lb]{\smash{\SetFigFont{5}{6.0}{rm}{}{}$\gamma$}}}
\put(1972,-2265){\makebox(0,0)[lb]{\smash{\SetFigFont{5}{6.0}{rm}{}{}$\gamma$}}}
\put(13635,-5869){\makebox(0,0)[lb]{\smash{\SetFigFont{5}{6.0}{rm}{}{}$R$}}}
\put(13092,-788){\makebox(0,0)[lb]{\smash{\SetFigFont{5}{6.0}{rm}{}{}$R'$}}}
\put(14839,-4999){\makebox(0,0)[lb]{\smash{\SetFigFont{5}{6.0}{rm}{}{}$\beta$}}}
\put(19505,-2199){\makebox(0,0)[lb]{\smash{\SetFigFont{5}{6.0}{rm}{}{}$\gamma$}}}
\put(19672,-4265){\makebox(0,0)[lb]{\smash{\SetFigFont{5}{6.0}{rm}{}{}$\gamma$}}}
\put(14172,-4849){\makebox(0,0)[lb]{\smash{\SetFigFont{5}{6.0}{rm}{}{}$\gamma$}}}
\put(12618,-70){\makebox(0,0)[lb]{\smash{\SetFigFont{5}{6.0}{rm}{}{}$x'$}}}
\put(21711,-1707){\makebox(0,0)[lb]{\smash{\SetFigFont{5}{6.0}{rm}{}{}$y'$}}}
\put(13230,-6528){\makebox(0,0)[lb]{\smash{\SetFigFont{5}{6.0}{rm}{}{}$x$}}}
\put(21900,-4555){\makebox(0,0)[lb]{\smash{\SetFigFont{5}{6.0}{rm}{}{}$y$}}}
\put(19054,-3681){\makebox(0,0)[lb]{\smash{\SetFigFont{5}{6.0}{rm}{}{}$y'_{j}$}}}
\put(19098,-2701){\makebox(0,0)[lb]{\smash{\SetFigFont{5}{6.0}{rm}{}{}$y_{j}$}}}
\put(15635,-3100){\makebox(0,0)[lb]{\smash{\SetFigFont{8}{9.6}{rm}{}{}$\Gamma_j$}}}
\put(14689,-4215){\makebox(0,0)[lb]{\smash{\SetFigFont{5}{6.0}{rm}{}{}$x_{j}$}}}
\put(13611,-2307){\makebox(0,0)[lb]{\smash{\SetFigFont{5}{6.0}{rm}{}{}$x'_{j}$}}}
\put(12605,-4165){\makebox(0,0)[lb]{\smash{\SetFigFont{5}{6.0}{rm}{}{}$\gamma$}}}
\put(12638,-2282){\makebox(0,0)[lb]{\smash{\SetFigFont{5}{6.0}{rm}{}{}$\gamma$}}}
\put(5407,-1237){\makebox(0,0)[lb]{\smash{\SetFigFont{5}{6.0}{rm}{}{}$\Gamma_j(x'_j,y'_j)$}}}
\put(5774,-5104){\makebox(0,0)[lb]{\smash{\SetFigFont{5}{6.0}{rm}{}{}$\Gamma_j(x_j,y_j)$}}}
\put(16241,-2187){\makebox(0,0)[lb]{\smash{\SetFigFont{5}{6.0}{rm}{}{}$\Gamma_j(x'_j,y'_j)$}}}
\end{picture}
\caption{The walks $R$ and $R'$ that intersect $\Gamma_j$}\label{fig:RetR'}
\end{center}
\end{figure}

\begin{Lem}\label{lem:Construction de_R_et_R'}
Let $G$ be a cubic bridgeless traceable graph such that
$\chi'(G)=4$. Let $\Gamma_j$ be an even cycle of the sequence
$(\Gamma_j)_{j=1\ldots h}$. Then there are two $\alpha$-disjoint
walks say $R$ and $R'$ such that
\begin{enumerate}
\item $R(x,x_j)=Q(x,x_j)$ and $R'(x',x'_j)=Q'(x',x'_j)$.
\item $R$ contains one end vertex of $e_{max(\Gamma_j)}$,  say
$y_{j}$, while $R'$ contains the other, say $y'_{j}$.
\item Either $Q(y_j,y)$ is a subwalk of $R$ and $Q'(y'_j,y')$ a subwalk of $R'$ or $Q'(y'_j,y')$ is
a subwalk of $R$ and $Q(y_j,y)$ a subwalk of $R'$.
\item $R(x_j,y_j)$ and $R'(x'_j,y'_j)$ are subpaths of $\Gamma_{j}$.
\item $R(x_j,y_j)$ is a $\gamma$-chain.
\end{enumerate}
\end{Lem}
\begin{Prf}
One of the two paths of $\Gamma_{j}$ joining $x_{j}$ to the endpoints of $e_{max(\Gamma_j)}$ is certainly a $\gamma$-chain. Let $P$ be this path. Let $y_j$ be the endpoint of $e_{max(\Gamma_j)}$ which belongs to $P$ while $y'_j$ denotes the other.
If $Q$ contains the path $P$, we set $R=Q$ and $R'=Q'$ otherwise we set
$R=Q(x,x_{j})+P+Q'(y_{j},y')$ and $R'=Q'(x',x'_{j})+P'+Q(y'_{j},y) $ where $P'$ is a subpath of $\Gamma_{j}$ joining $x'_{j}$ to $y'_j$ (see Figure \ref{fig:RetR'}).
\end{Prf}

In the following, up to a renaming of the vertices
$y$ and $y'$, we assume that $y$ is an endpoint of $R$ while $y'$ is
an endpoint of $R'$.

Since $\Gamma_j$ is an even cycle we have $j<h$, thus there certainly exists one cycle in
$\{\Gamma_{j+1},\Gamma_{j+2}\}$ say $\Gamma$ which have an endpoint of $e_{max(\Gamma)}$ on $R$.
The index $\Gamma$ in the sequence $(\Gamma_j)_{j=1\ldots h}$ will be denoted $\sigma_R(\Gamma_j)$.
The index $\sigma_{R'}(\Gamma_j)$ is defined similarly from the walk $R'$. By construction we have
$\{\sigma_{R}(\Gamma_j),\sigma_{R'}(\Gamma_j)\}=\{j+1,j+2\}$.

\begin{Lem}\label{Lem:UnWalkQuiWellIntersectsGammaJ}
Let $G$ be a cubic bridgeless traceable graph such that $\chi'(G)=4$. Let $\Gamma_j$ be an
even cycle of the sequence $(\Gamma_j)_{j=1\ldots h}$.
There are two $\alpha$-disjoint walks say $S$ and $S'$ such that
\begin{enumerate}
\item The vertex $x_j$ (resp. $x'_j$) is an endpoint of $S$ (resp. $S'$).
\item $S$ and $S'$ have distinct endpoints in $\{x_{\sigma_R(\Gamma_j)},x'_{\sigma_{R'}(\Gamma_j)}\}$.
\item The vertices of $S$ and $S'$ are vertices of $R(x_j,x_{\sigma_R(\Gamma_j)})$ or vertices of
$R'(x'_j,x_{\sigma_{R'}(\Gamma_j)})$ or of $\Gamma_j$.
\item $S$ well-intersects the cycle $\Gamma_j$.
\end{enumerate}
\end{Lem}
\begin{Prf}
By construction the walk $R(x_j,y_j)$ well-intersects $\Gamma_j$. If
the subwalk $R(y_i,\sigma_R(\Gamma_j))$ shares an edge, say $e$,
with $\Gamma_j$, this edge is coloured with $\beta$. When $e$
belongs  to $\Gamma(x_j,y_j)$ the intersection of $R$ with
$\Gamma_j$ will not be changed by $e$. This is not the case when $e$ is an edge of a
$\beta$-chain.

Let $P$ be the subpath of $\Gamma_j$ whose endpoints are $x_j$
and $y_j$ which is distinct from $\Gamma_j(x_j,y_j)$. Observe that
$P$ is a $\beta$-chain.

If the subwalk $R(y_i,\sigma_R(\Gamma_j))$ does not intersect $P$  we set $S=R(y_i,\sigma_R(\Gamma_j))$ and
$S'=R'(x'_j,x'_{\sigma_{R'}(\Gamma_j)})$ and we are done.

If on the contrary $R(y_i,\sigma_R(\Gamma_j))$ shares an edge with
$P$, let $ab$ ($a\prec_R b$) be
a such edge where $\Gamma(a,y_j)$ is a subpath of $P$ with maximum length. It must be pointed out that in this case
$R(y_i,\sigma_R(\Gamma_j))$ does not intersect with $\Gamma_j$.

{\bf Case $1$.} If $b$ is a vertex of $\Gamma(a,y_j)$ (see Figure
\ref{fig:Cas1}) we write
$S=R(x_j,y_j)+\Gamma(y_j,b)+R(b,\sigma_R(\Gamma_j))$ and
$S'=R'(x'_j,\sigma_R'(\Gamma_j))$ .
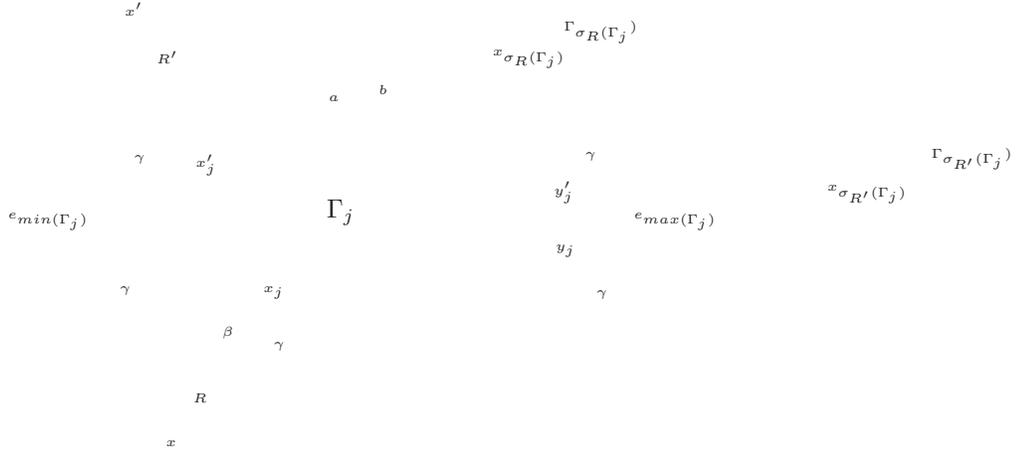
\begin{figure}
\begin{center}
\begin{picture}(0,0)%
\special{psfile=Cas1.ps}%
\end{picture}%
\setlength{\unitlength}{1657sp}%
\begingroup\makeatletter\ifx\SetFigFont\undefined
\def\x#1#2#3#4#5#6#7\relax{\def\x{#1#2#3#4#5#6}}%
\expandafter\x\fmtname xxxxxx\relax \def\y{splain}%
\ifx\x\y   
\gdef\SetFigFont#1#2#3{%
  \ifnum #1<17\tiny\else \ifnum #1<20\small\else
  \ifnum #1<24\normalsize\else \ifnum #1<29\large\else
  \ifnum #1<34\Large\else \ifnum #1<41\LARGE\else
     \huge\fi\fi\fi\fi\fi\fi
  \csname #3\endcsname}%
\else
\gdef\SetFigFont#1#2#3{\begingroup
  \count@#1\relax \ifnum 25<\count@\count@25\fi
  \def\x{\endgroup\@setsize\SetFigFont{#2pt}}%
  \expandafter\x
    \csname \romannumeral\the\count@ pt\expandafter\endcsname
    \csname @\romannumeral\the\count@ pt\endcsname
  \csname #3\endcsname}%
\fi
\fi\endgroup
\begin{picture}(15013,7656)(84,-6605)
\put(13900,-2209){\makebox(0,0)[lb]{\smash{\SetFigFont{5}{6.0}{rm}{}{}$\Gamma_{\sigma_{R'}(\Gamma_j})$}}}
\put(9451,-3121){\makebox(0,0)[lb]{\smash{\SetFigFont{5}{6.0}{rm}{}{}$e_{max(\Gamma_j)}$}}}
\put( 84,-3121){\makebox(0,0)[lb]{\smash{\SetFigFont{5}{6.0}{rm}{}{}$e_{min(\Gamma_j)}$}}}
\put(7334,-686){\makebox(0,0)[lb]{\smash{\SetFigFont{5}{6.0}{rm}{}{}$x_{\sigma_R(\Gamma_j)}$}}}
\put(12344,-2713){\makebox(0,0)[lb]{\smash{\SetFigFont{5}{6.0}{rm}{}{}$x_{\sigma_{R'}(\Gamma_j)}$}}}
\put(1972,-2265){\makebox(0,0)[lb]{\smash{\SetFigFont{5}{6.0}{rm}{}{}$\gamma$}}}
\put(1756,-4231){\makebox(0,0)[lb]{\smash{\SetFigFont{5}{6.0}{rm}{}{}$\gamma$}}}
\put(2852,-5886){\makebox(0,0)[lb]{\smash{\SetFigFont{5}{6.0}{rm}{}{}$R$}}}
\put(2309,-805){\makebox(0,0)[lb]{\smash{\SetFigFont{5}{6.0}{rm}{}{}$R'$}}}
\put(3289,-4882){\makebox(0,0)[lb]{\smash{\SetFigFont{5}{6.0}{rm}{}{}$\beta$}}}
\put(8722,-2215){\makebox(0,0)[lb]{\smash{\SetFigFont{5}{6.0}{rm}{}{}$\gamma$}}}
\put(8889,-4282){\makebox(0,0)[lb]{\smash{\SetFigFont{5}{6.0}{rm}{}{}$\gamma$}}}
\put(4056,-5065){\makebox(0,0)[lb]{\smash{\SetFigFont{5}{6.0}{rm}{}{}$\gamma$}}}
\put(1835,-87){\makebox(0,0)[lb]{\smash{\SetFigFont{5}{6.0}{rm}{}{}$x'$}}}
\put(2447,-6545){\makebox(0,0)[lb]{\smash{\SetFigFont{5}{6.0}{rm}{}{}$x$}}}
\put(8254,-2764){\makebox(0,0)[lb]{\smash{\SetFigFont{5}{6.0}{rm}{}{}$y'_{j}$}}}
\put(8281,-3601){\makebox(0,0)[lb]{\smash{\SetFigFont{5}{6.0}{rm}{}{}$y_{j}$}}}
\put(4852,-3117){\makebox(0,0)[lb]{\smash{\SetFigFont{10}{12.0}{rm}{}{}$\Gamma_j$}}}
\put(3906,-4232){\makebox(0,0)[lb]{\smash{\SetFigFont{5}{6.0}{rm}{}{}$x_{j}$}}}
\put(2894,-2356){\makebox(0,0)[lb]{\smash{\SetFigFont{5}{6.0}{rm}{}{}$x'_{j}$}}}
\put(4883,-1361){\makebox(0,0)[lb]{\smash{\SetFigFont{5}{6.0}{rm}{}{}$a$}}}
\put(5634,-1267){\makebox(0,0)[lb]{\smash{\SetFigFont{5}{6.0}{rm}{}{}$b$}}}
\put(8402,-315){\makebox(0,0)[lb]{\smash{\SetFigFont{5}{6.0}{rm}{}{}$\Gamma_{\sigma_R(\Gamma_j})$}}}
\end{picture}
\caption{The walks $R$ and $R'$ in Case $1$}\label{fig:Cas1}
\end{center}
\end{figure}

{\bf Case $2$.} When $b$ does not belong to  $\Gamma(a,y_j)$ the $a$ is a
vertex of $\Gamma(b,y_j)$( see Figure \ref{fig:Cas2}). We write
$S=R(x_j,y_j)+R(y_j,a)+\Gamma(a,y'_j)+R'(y'_j,\sigma_{R'}(\Gamma_j))$
and $S'=R'(x'_j,b)+R(b,\sigma_R(\Gamma_j))$, where $R'(x'_j,b)$
denotes the subpath of $\Gamma$ with endpoints $x'_j$ and $b$ which
does not contain $a$.

\begin{figure}
\begin{center}
\begin{picture}(0,0)%
\special{psfile=Cas2.ps}%
\end{picture}%
\setlength{\unitlength}{1657sp}%
\begingroup\makeatletter\ifx\SetFigFont\undefined
\def\x#1#2#3#4#5#6#7\relax{\def\x{#1#2#3#4#5#6}}%
\expandafter\x\fmtname xxxxxx\relax \def\y{splain}%
\ifx\x\y   
\gdef\SetFigFont#1#2#3{%
  \ifnum #1<17\tiny\else \ifnum #1<20\small\else
  \ifnum #1<24\normalsize\else \ifnum #1<29\large\else
  \ifnum #1<34\Large\else \ifnum #1<41\LARGE\else
     \huge\fi\fi\fi\fi\fi\fi
  \csname #3\endcsname}%
\else
\gdef\SetFigFont#1#2#3{\begingroup
  \count@#1\relax \ifnum 25<\count@\count@25\fi
  \def\x{\endgroup\@setsize\SetFigFont{#2pt}}%
  \expandafter\x
    \csname \romannumeral\the\count@ pt\expandafter\endcsname
    \csname @\romannumeral\the\count@ pt\endcsname
  \csname #3\endcsname}%
\fi
\fi\endgroup
\begin{picture}(14673,7656)(217,-6605)
\put(12344,-2713){\makebox(0,0)[lb]{\smash{\SetFigFont{5}{6.0}{rm}{}{}$x_{\sigma_{R'}(\Gamma_j)}$}}}
\put(9451,-3121){\makebox(0,0)[lb]{\smash{\SetFigFont{5}{6.0}{rm}{}{}$e_{max(\Gamma_j)}$}}}
\put(217,-3088){\makebox(0,0)[lb]{\smash{\SetFigFont{5}{6.0}{rm}{}{}$e_{min(\Gamma_j)}$}}}
\put(13582,-2191){\makebox(0,0)[lb]{\smash{\SetFigFont{5}{6.0}{rm}{}{}$\Gamma_{\sigma_{R'}(\Gamma_j})$}}}
\put(8328,-352){\makebox(0,0)[lb]{\smash{\SetFigFont{5}{6.0}{rm}{}{}$\Gamma_{\sigma_R(\Gamma_j})$}}}
\put(4217,-1494){\makebox(0,0)[lb]{\smash{\SetFigFont{5}{6.0}{rm}{}{}$b$}}}
\put(4883,-1361){\makebox(0,0)[lb]{\smash{\SetFigFont{5}{6.0}{rm}{}{}$a$}}}
\put(2894,-2356){\makebox(0,0)[lb]{\smash{\SetFigFont{5}{6.0}{rm}{}{}$x'_{j}$}}}
\put(3906,-4232){\makebox(0,0)[lb]{\smash{\SetFigFont{5}{6.0}{rm}{}{}$x_{j}$}}}
\put(4852,-3117){\makebox(0,0)[lb]{\smash{\SetFigFont{10}{12.0}{rm}{}{}$\Gamma_j$}}}
\put(8281,-3601){\makebox(0,0)[lb]{\smash{\SetFigFont{5}{6.0}{rm}{}{}$y_{j}$}}}
\put(8254,-2764){\makebox(0,0)[lb]{\smash{\SetFigFont{5}{6.0}{rm}{}{}$y'_{j}$}}}
\put(2447,-6545){\makebox(0,0)[lb]{\smash{\SetFigFont{5}{6.0}{rm}{}{}$x$}}}
\put(1835,-87){\makebox(0,0)[lb]{\smash{\SetFigFont{5}{6.0}{rm}{}{}$x'$}}}
\put(4056,-5065){\makebox(0,0)[lb]{\smash{\SetFigFont{5}{6.0}{rm}{}{}$\gamma$}}}
\put(8889,-4282){\makebox(0,0)[lb]{\smash{\SetFigFont{5}{6.0}{rm}{}{}$\gamma$}}}
\put(8722,-2215){\makebox(0,0)[lb]{\smash{\SetFigFont{5}{6.0}{rm}{}{}$\gamma$}}}
\put(3289,-4882){\makebox(0,0)[lb]{\smash{\SetFigFont{5}{6.0}{rm}{}{}$\beta$}}}
\put(2309,-805){\makebox(0,0)[lb]{\smash{\SetFigFont{5}{6.0}{rm}{}{}$R'$}}}
\put(2852,-5886){\makebox(0,0)[lb]{\smash{\SetFigFont{5}{6.0}{rm}{}{}$R$}}}
\put(1756,-4231){\makebox(0,0)[lb]{\smash{\SetFigFont{5}{6.0}{rm}{}{}$\gamma$}}}
\put(1972,-2265){\makebox(0,0)[lb]{\smash{\SetFigFont{5}{6.0}{rm}{}{}$\gamma$}}}
\put(7184,-649){\makebox(0,0)[lb]{\smash{\SetFigFont{5}{6.0}{rm}{}{}$x_{\sigma_R(\Gamma_j)}$}}}
\end{picture}
\caption{The walks $R$ and $R'$ in Case $2$ }\label{fig:Cas2}
\end{center}
\end{figure}
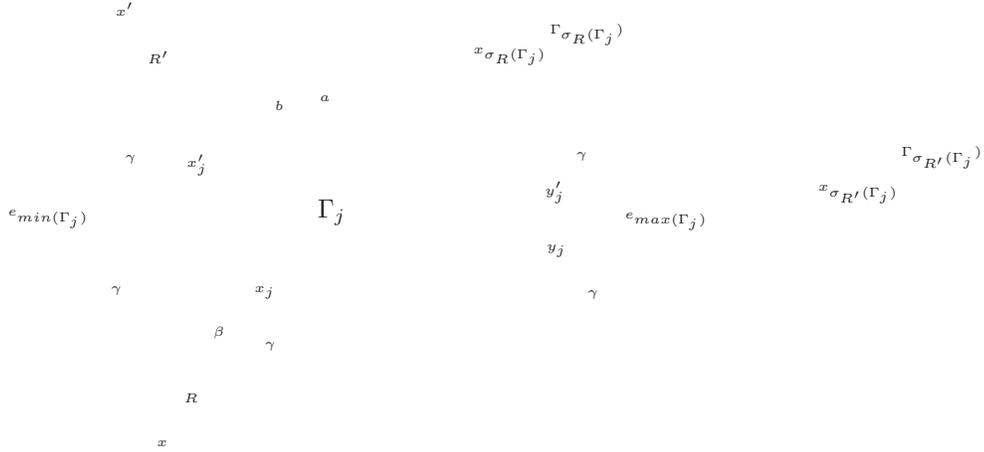
\end{Prf}
\subsection{The main results}
We use in the sequel the same notations than above.

In Propositions \ref{prop:FanRaspaudQuandOna2Cycles}, \ref{prop:FanRaspaudQuandOna3Cycles} and \ref{prop:FanRaspaudQuandLesCyclesSontSansCroisement} we consider particular cases of cubic graph for which Conjecture \ref{Conjecture:FanRaspaud} holds true.
\begin{Prop}\label{prop:FanRaspaudQuandOna2Cycles}
Let $G$ be a cubic bridgeless traceable graph such that
$\chi'(G)=4$. If the sequence $(\Gamma_j)_{j=1\ldots h}$ has only
two cycles then there exists four perfect matchings $M_{\alpha}$,
$M_1$, $M_2$ and $M_3$ such that $M_{\alpha}\cap M_i\cap
M_j=\emptyset$ for $i,j\in \{1,2,3\}$, $i\neq j$.
\end{Prop}

\begin{Prf}
Since $h=2$ we have $\Gamma_1=C_1$ and $\Gamma_2=C_k$. Moreover, the
walks described in Lemma \ref{Lem:TroisWalksDeC1ACn} are reduced to
paths whose edges are alternately coloured with $\alpha$ and $\beta$. Thus, for $i\in \{1,2,3\}$, by
Lemma \ref{Lem:CouplageEquilibréG_M-V(A)}, the set of $\alpha$-edges of $W_i$ is a balanced $M_{\alpha}-$matching. Hence we are done since the walks $W_1$, $W_2$ and
$W_3$ are $\alpha$-disjoint.
\end{Prf}

\begin{Prop}\label{prop:FanRaspaudQuandOna3Cycles}
Let $G$ be a cubic bridgeless traceable graph such that
$\chi'(G)=4$. If the sequence $(\Gamma_j)_{j=1\ldots h}$ has only
three cycles then there exists three perfect matchings $M_{\alpha}$,
$M_1$, $M_2$  such that $M_{\alpha}\cap M_1\cap M_2=\emptyset$
\end{Prop}

\begin{Prf}
As a matter of fact, when $h=3$, the sequence $(\Gamma_j)_{j=1\ldots 3}$ is reduced to  $(\Gamma_{1}=C_{1}, \Gamma_{2}, \Gamma_{3}=C_{k})$.
Let $W_{1}, W_{2}$ and $W_{3}$ be the three walks obtained by Lemma
\ref{Lem:TroisWalksDeC1ACn}. By Lemma \ref{Lemma:WellIntersectedNotInvolved} those walks well-intersect all cycles which are not in the sequence $(\Gamma_j)_{j=1\ldots h}$.
By Lemma \ref{Lemma:WellIntersectedGammajNotInvolved} we can consider that $W_{2}$
well-intersects $\Gamma_{2}$ and by Lemma \ref{Lem:UnWalkQuiWellIntersectsGammaJ} we can transform $W_{1}$ in a walk $S_{1}$ well-intersecting this cycle. We get hence two walks $\alpha$ disjoint $S_{1}$ and $W_{2}$ well-intersecting every cycle
of $G_{M_{\alpha}}$.
Hence the set of edges of $W$ coloured with $\alpha$, as well as the same set for $W'$, are balanced  $M_{\alpha}-$matchings.  By Lemma \ref{Lem:CouplageEquilibréG_M-V(A)}, we get thus two perfect matchings $M_{1}$ and $M_{2}$ such that $M_{\alpha}\cap M_1\cap M_2=\emptyset$ as claimed.
\end{Prf}

Let $G$ be a cubic bridgeless traceable graph such that $\chi'(G)=4$.
Recall that given an even  cycle
of the sequence $(\Gamma_j)_{j=1\ldots h}$ say $\Gamma_j$ we have defined two odd paths $P_j$ and $P'_j$ with one end in $\{min(\Gamma_j), min(\Gamma_j)+1\}$ and the other end in
$\{max(\Gamma_j), max(\Gamma_j)+1\}$. When one of those paths have endpoints in $\{min(\Gamma_j), max(\Gamma_j)\}$ the paths $P_j$ and $P'_j$ are said to be {\em crossing }, {\em non crossing} otherwise.

\begin{Prop}\label{prop:FanRaspaudQuandLesCyclesSontSansCroisement}
Let $G$ be a cubic bridgeless traceable graph such that
$\chi'(G)=4$. If for each even  cycle of the sequence
$(\Gamma_j)_{j=1\ldots h}$, say $\Gamma_j$, the paths $P_j$ and $P'_j$ are  non crossing then there exists
three perfect matchings $M_{\alpha}$, $M_1$, $M_2$  such that
$M_{\alpha}\cap M_1\cap M_2=\emptyset$
\end{Prop}
\begin{Prf}
By Propositions \ref{prop:FanRaspaudQuandOna2Cycles} and
\ref{prop:FanRaspaudQuandOna3Cycles}, we can consider that
$h \geq 4$. Let $W_{1}, W_{2}$ and $W_{3}$ be the $3$ walks
obtained by Lemma \ref{Lem:TroisWalksDeC1ACn}. It is an easy task to
see that, up to the names of the walks, $W_{1}$ is obtained by
replacing the additional edges with the paths that they represent for
the even cycles $\Gamma_{j}$ with $j$ even. In the same way, $W_{2}$
is obtained by replacing the additional edges with the paths that they
represent for the even cycles  $\Gamma_{j}$ with $j$ odd. At last,
$W_{3}$ is obtained by replacing the additional edges with the paths
that they represent for each even cycles $\Gamma_{j}$.

Starting with the  {\em green} colour for the subpath containing the
vertex $1$, the maximal subpaths of the Hamiltonian path, say $P$, not containing the edges
$e_{min(\Gamma_{j})}$ and $e_{max(\Gamma_{j})}$ ($j=2\ldots h-1$),
as well as the edges $e_{max(\Gamma_{1})}$ and
$e_{min(\Gamma_{h})}$, are
coloured alternately with {\em green} and {\em red}. One can see
that $W_{3}$ uses all the red subpaths while $W_{1}$ and $W_{2}$
use the green subpaths only.
\begin{ClaimSansNum}\label{Claim:W1WellIntersect}
$W_{i}$ ($i=1,2$) well intersects each cycle of the sequence
$(\Gamma_j)_{j=1\ldots h}$
\end{ClaimSansNum}
\begin{PrfClaim}
Assume without loss of generality that $i=1$. From Lemmas \ref{Lemma:WellIntersectedNotInvolved} and
\ref{Lemma:WellIntersectedGammajNotInvolved}
we have just to prove that
$W_{1}$ well intersects the even cycles $\Gamma_{j}$ with $j\geq 2$
even.

We can check that $W_{1}$ contains the vertex $min(\Gamma_j)$, moreover since the subpaths $P_j$ and $P'_j$ of $\Gamma_j$ are not crossing the vertex $max(\Gamma_j)+1$ belongs to $W_1$ too.

By construction of $\Gamma_{j}$, the green subpath of $P$ ending with the vertex $min(\Gamma_{j})$ has no edge in common with $\Gamma_{j}$, as well as the green subpath starting with $max(\Gamma_{j})+1$. Hence $W_{1}$ contains exactly two $\alpha$-edges  one ending on $min(\Gamma_{j})$
and the other on  $max(\Gamma_{j})+1$ on $\Gamma_{j}$. The two subpaths of
$\Gamma_{j}$ so determined are odd, which proves that $W_{1}$ well
intersect $\Gamma_{j}$.
\end{PrfClaim}

By Lemma
\ref{Lem:CouplageEquilibréG_M-V(A)}, the set of $\alpha$-edges of
$W_1$ ($W_{2}$ respectively) is a balanced $M_{\alpha}$-matching. We
have thus two perfect matchings $M_{1}$ and $M_{2}$ with no $\alpha$-edge in common. Hence $M_{\alpha}, M_{1}$ and $M_{2}$ have an empty intersection, as claimed.
\end{Prf}

Conjecture \ref{Conjecture:KaiserRaspaud} is known to be verified for
bridgeless cubic graphs of oddness $2$ (see \cite{KaiRas}), Theorem
\ref{Thm:CubicTraceable} gives a stronger result for cubic
bridgeless traceable graphs of chromatic index $4$.

\begin{Thm}\label{Thm:CubicTraceable}
Let $G$ be a cubic bridgeless traceable graph of chromatic index
$4$. Then there exists four perfect matchings $M_{\alpha}$, $M_1$,
$M_2$ and $M_3$ such that $M_{\alpha}\cap M_i$ does not contain any
odd cut set, for $i\in\{1,2,3\}$.
Moreover, for $i\in\{1,2,3\}$ one can associate to $M_i$ two joins
$J_i$ and $J'_i$ such that $M_{\alpha}\cap M_i\cap
J_i=M_{\alpha}\cap M_i\cap J'_i=\emptyset$.
\end{Thm}
\begin{Prf}
We consider the walks $W_1$, $W_2$ and $W_3$ described in Lemma
\ref{Lem:TroisWalksDeC1ACn}. Without loss of generality we choose
$i=1$ and we derive from $W_1$ a walk $S_1$ as follows.

First we set $S_1=W_1$. Following the
natural order on the vertices of $S_1$ given by $\prec_{S_1}$ when
there is on $S_1$ a vertex of some edge $e_{max(\Gamma_j)}$ for some
cycle $\Gamma_j$ of the sequence $(\Gamma_j)_{j=2,\ldots h-1}$ we
set $R=S_1$ and thus either $R'=W_2$ or $R'=W_3$. We apply Lemma
\ref{Lem:UnWalkQuiWellIntersectsGammaJ} on $R$ and $R'$ and we get thus two walks $S$ and
$S'$, we know that $R(x,x'_j)+S$ well-intersects $\Gamma_j$. The walks $S$ and $S'$ have endpoints in $\{\sigma_R(\Gamma_j),\sigma_{R'}(\Gamma_j)\}$. Hence when $\sigma_R(\Gamma_j)$ is an endpoint of $S$ we write $S_1=S_1(x,x_j)+S+R(\sigma_R(\Gamma_j),y)$ and $R'=R'(x',x'_j)+S'+R'(\sigma_{R'}(\Gamma_j),y')$. Otherwise $\sigma_{R'}(\Gamma_j)$ is an endpoint of $S$ and we write $S_1=S_1(x,x'_j)+S+R'(\sigma_{R'}(\Gamma_j),y')$ and $R'=R'(x',x'_j)+S'+R(\sigma_{R}(\Gamma_j),y)$(recall that either $R'=W_2$ or $R'=W_3$).
Finally $S_1$ well-intersects all concerned cycle of the sequence $(\Gamma_j)_{j=2,\ldots h-1}$. Moreover,
the walks $S_1$, $W_2$ and $W_3$ are $\alpha$-disjoint all of them have one endpoint in $\{q_1,q_2,q_3\}$ and the
other one in $\{q'_1,q'_2,q'_3\}$.

Due to Lemma \ref{Lem:CouplageEquilibréG_M-V(A)}, the set of edges
$A=M_{\alpha}\cap S_1$ is a balanced $M_{\alpha}$-matching, that is
there is a perfect matching $M_1$ such that $M_{\alpha}\cap M_1=A$.
\newline But now, if $M_{\alpha}\cap M_1$ contains an odd cut set,
say $X$, there must be a partition $(V_1,V_2)$ of $V(G)$ into two
odd sets such that the edges of $X$ have one end in $V_1$ and the
other end in $V_2$. Moreover $X\subset M_{\alpha}$, $V_1$ and $V_2$ being odd, each
of those sets precisely contains exactly one odd cycle of $G_{M_{\alpha}}$. Since $W_2$ and $W_3$
are both connecting a vertex of $C_1$ to a vertex of $C_k$, there must be an edge of $S_1$ and an edge
of $W_3$ in $X$, a contradiction since $S_1$, $W_2$ and $W_3$ are $\alpha$-disjoint.

Moreover, the set of vertices $S_1\cup W_2- S_1\cap W_2$ together
with a sub-path of $C_1$ whose endpoints are $q_1$ and $q_2$ and a
sub-path of $C_k$ whose endpoints are $q'_1$ and $q'_2$ form a set
$X$ of vertices that induce cycles of $G$. Thus the
 edge set $J_1$ of the subgraph induced
with $V(G)-X$ is a join that avoids the edges of $S_1$, in other
words $M_{\alpha}\cap M_1\cap J_1=\emptyset$. Similarly we can derive
from $S_1\cup W_3-S_1\cap W_3$ another join $J'_1$ with the same
property.
\end{Prf}

A direct consequence of Theorem \ref{Thm:CubicTraceable} is that Conjecture \ref{Conj:MakajovaSkoviera} holds true for cubic bridgeless traceable graphs.

\section{Conclusion}
As far as we know the techniques developed in Lemmas \ref{Lem:TroisWalksDeC1ACn} to  \ref{Lem:UnWalkQuiWellIntersectsGammaJ} as well as in Theorem \ref{Thm:CubicTraceable} do not lead  to a proof of Conjecture \ref{Conjecture:FanRaspaud} for cubic bridgeless traceable graphs in general. However Conjecture \ref{Conjecture:FanRaspaud} holds true in some particular cases, see for example Propositions \ref{prop:FanRaspaudQuandOna2Cycles}, \ref{prop:FanRaspaudQuandOna3Cycles} or \ref{prop:FanRaspaudQuandLesCyclesSontSansCroisement} or when in applying Lemma \ref{Lem:UnWalkQuiWellIntersectsGammaJ} on all the concerned cycles of the sequence $(\Gamma_j)_{j=2,\ldots h-1}$, we get two $\alpha$-disjoint walks that well-intersect the cycles.

In a forthcoming paper (\cite{FouVan08}), we prove that a minimal counter-example to Conjecture
\ref{Conjecture:FanRaspaud} must have at least $36$ vertices ($40$ vertices when the cyclic edge connectivity of the graph is at least $4$).

{\bf Acknowledgment.} The authors are grateful to the anonymous referees for their helpful corrections and comments.

\bibliographystyle{plain}
\bibliography{BibliographieTraceable}
\end{document}